\pdfoutput=1
\documentclass{article}
\usepackage{amsmath}
\usepackage{arxiv}
\usepackage[utf8]{inputenc} 
\usepackage[T1]{fontenc}    
\usepackage{color}
\usepackage[unicode=true,pdfusetitle,
 bookmarks=true,bookmarksnumbered=false,bookmarksopen=false,
 breaklinks=true,pdfborder={0 0 0},pdfborderstyle={},backref=false,colorlinks=true]
 {hyperref}
\hypersetup{
 linkcolor=red, urlcolor=magenta, citecolor=blue}

\usepackage{url}            
\usepackage{booktabs}       
\usepackage{amsfonts}       
\usepackage{nicefrac}       
\usepackage{microtype}      
\usepackage{lipsum}		
\usepackage{graphicx}
\usepackage{doi}
\usepackage{orcidlink}
\usepackage{array}
\usepackage[style=authoryear,backend=bibtex]{biblatex}

\title{CogNarr Ecosystem: Facilitating Group Cognition at Scale}

\author{John C. Boik\,\orcidlink{0000-0003-1289-7997} \\
	Active Inference Institute\\
	Davis, CA USA\\
	\texttt{john.boik@activeinference.institute} \\
}

\date{}

\renewcommand{\shorttitle}{CogNarr Ecosystem: Facilitating Group Cognition at Scale}

\begin{document}

\maketitle

\begin{abstract}
Human groups of all sizes and kinds engage in deliberation, problem
solving, strategizing, decision making, and more generally, cognition.
Some groups are large, and that setting presents unique challenges.
The small-group setting often involves face-to-face dialogue, but
group cognition in the large-group setting typically requires some
form of online interaction. New approaches are needed to facilitate
the kind of rich communication and information processing that are
required for effective, functional cognition in the online setting,
especially for groups characterized by thousands to millions of participants
who wish to share potentially complex, nuanced, and dynamic perspectives.
This concept paper proposes the CogNarr (Cognitive Narrative) ecosystem,
which is designed to facilitate functional cognition in the large-group
setting. The paper's contribution is a novel vision as to how recent
developments in cognitive science, artificial intelligence, natural
language processing, and related fields might be scaled and applied
to large-group cognition, using an approach that itself promotes further
scientific advancement. A key perspective is to view a group as an
organism that uses some form of cognitive architecture to sense the
world, process information, remember, learn, predict, make decisions,
and adapt to changing conditions. The CogNarr ecosystem is designed
to serve as a component within that architecture. 
\end{abstract}

\maketitle

\section{Introduction}

Consider a small group of people in a room who wish to understand
each other's perspectives on some situation or problem. The group
might also want to deliberate and collaboratively develop and decide
upon a favored course of action. If discourse is reasonably healthy
(e.g., honest and respectful), this type of small-group setting can
allow participants to share rich, even nuanced and complex information
about their beliefs, desires, intentions, concerns, and observations.
In turn, rich communication can facilitate effective decision making
\parencite{aghazadehardebiliLiteratureReview2020}. But how can rich
information be shared in a group that numbers in the hundreds, thousands,
or even millions of participants? In the large-group setting, which
typically involves online interaction, how can each person's potentially
complex, nuanced, and dynamic perspective be heard and understood
by the group as a whole? The CogNarr (Cognitive Narrative) ecosystem,
proposed here, seeks to answer these questions.

Participants of a group engage in a cognitive process that can involve
sensing, deliberation, and decision making. One can attribute that
process to individual group members, but also to the group as a whole.
The distinction between cognition in individuals and in groups is
not as crisp as once believed \parencite{gliksonIntelligentCommunities2019}.
Lyon and colleagues list thirteen hallmarks of basal cognition, including
sensing, remembering, learning, anticipating (predicting), and problem
solving \parencite{lyonCognitiveCell2015,lyonReframingCognition2021}.
These hallmarks, listed later in full, are not unique to humans---all
thirteen appear in individuals across species, even in plants and
unicellular organisms. And they are not unique to individuals---they
appear also in groups across species. \textcite{watsonCollectiveIntelligence2023}
suggest that an organism can be defined as a collective of cooperating
entities that displays a cognitive architecture. The collective might
consist of cells, bees, or humans, for example. The organism's cognitive
architecture integrates information and coordinates action---the
core functionalities of cognition---regardless of whether it is implemented
by ``chemical, gene-regulatory, bio-electrical, neural, ecological,
or social interactions.'' Although not specifically mentioned by
the authors, we can add ``electrical'' to the list.

Thus, a coordinated group of humans in the CogNarr setting can be
understood as a cognitive organism that employs a cognitive architecture.
Moreover, CogNarr itself can be viewed as a component of that architecture.
Experience tells us that the cognitive process of any particular group
can range from highly functional to highly dysfunctional. The question
at hand, then, is what CogNarr design would favor the former and discourage
the latter? 

This concept paper contributes a novel vision as to how recent scientific
developments in cognitive science, artificial intelligence, natural
language processing, and related fields might be scaled and applied
to large-group cognition, using an approach that itself promotes further
scientific advancement. That vision, the CogNarr ecosystem, is still
in its incubation phase and the design of CogNarr is still incomplete.
The vision is presented here as a first, introductory step forward.
The CogNarr ecosystem is conceived to be an extensible set of technologies,
resources, applications (apps), tools, libraries, and services that
facilitates group cognition, especially for large groups, and especially
with regard to deliberation, strategizing, collaborative problem solving,
and decision making. The current paper provides a high-level overview
of the CogNarr ecosystem and a companion paper \parencite{boikCogNarrEcosystem2024b}
dives deeper into technical aspects of meaning representations for
user input. 

There is need for this project. While humans today can communicate
with potentially millions of others via social media platforms, those
platforms were not designed to support functional group cognition,
at least not in the sense described here. If the purpose of a tool
or platform is to facilitate group cognition, then each of Lyon's
thirteen hallmarks of cognition quite reasonably becomes a driver
of design. So too do topics such as transparency, power-sharing, privacy
protection, inclusiveness, governance, and information quality. Apart
from social media, several online tools and platforms have been developed
in recent years to facilitate group deliberation and decision making.
Some are discussed in Section 9, Related Work. While these can be
useful, there is still great room and need for more comprehensive
designs, such as CogNarr's, that better reflect our science-based
understanding of cognition and that are more capable of processing
the volume and richness of information that is natural in the large-group
setting. 

This project is timely. Societies are under stress due to a host of
unsolved or inadequately solved problems, including climate change,
biodiversity loss, pollution, economic and political instabilities,
and wealth and power inequalities, including poverty. These and other
problems exacerbate human suffering and can lead to social unrest.
Indeed, the very existence of multiple, serious, unsolved or inadequately
solved problems might be a symptom of long-standing dysfunctional
group cognition. Given that each unsolved problem can have local,
regional, national, and global impacts, groups of all sizes and kinds
are challenged. New tools that facilitate functional group cognition
might increase our capacity to address and solve problems, prevent
and recover from harm, and identify and act on shared purpose. 

Despite its incubation status, the CogNarr ecosystem is often described
here as if it already exists. This literary device avoids some repetition
of phrases such as ``could be'' or ``would be''. The intent, however,
is not to unnecessarily fix or limit the ecosystem design. As the
project develops, that design would likely change to some degree from
what is presented here. Importantly, the ecosystem could exist. That
is, the project is practical in the sense that a functional proof
of concept and minimum viable product could be developed using existing
technologies. A fully functional CogNarr ecosystem would benefit from
new scientific advances and emerging technologies. 

To further the scientific effort, to help ensure that the technology
is put to safe and best use, and to maximize transparency and build
and maintain trust within a population of users, the intention is
to develop and implement the ecosystem within an open source framework,
under an appropriate governance system. Third-party enhancements resting
on top of the core stack, and second- and third-party provision of
services (such as consulting and training) are possible. 

The CogNarr concept is a continuation of ideas presented in \textcite{boikEconomicDirect2014,boikScienceDrivenSocietal2020,boikScienceDrivenSocietal2020a,boikScienceDrivenSocietal2021},
which as a body of work addresses two questions: (1) Out of all conceivable
designs for economic, financial, governance, legal, educational, and
other core societal systems, which designs might best serve the common
good? and (2) By which viable strategies might new designs be developed,
tested, implemented, and monitored? In posing these questions, societal
systems are viewed as central components of a society's cognitive
architecture. The connection to the present work is that here the
concept of cognitive architecture is applied to the more modest but
still ambitious goal of developing a tool to facilitate group cognition,
where the group does not necessarily implement a suite of societal
systems. The same principles apply, however, and central questions
remain the same. Which designs for group cognitive architectures are
fit for purpose, and how can we best implement new and better designs? 

The CogNarr project is in the spirit of \textit{ecosystems of intelligence},
as envisioned by \textcite{fristonDesigningEcosystems2022}, although
there are major differences between that work and this one. While
the idea of shared narratives reflecting shared generative models
(of beliefs) is central to both, this paper takes the concept literally---sharing
belief models, conceived of as narratives, about specific problems
or situations, within specific groups. In contrast, Friston et al.
appear to aim toward a domain-general collective intelligence that
exists as a global web of generative models and intelligent systems,
with humans in the middle. Even in this difference, however, there
is potential complement. If the CogNarr ecosystem were to be constructed
and widely used, it would eventually generate a massive collection
of shared narratives that span a multitude of topics and domains.
That repository, and the continuously improving CogNarr computational
system at its core, might one day constitute a node in the network
of intelligence envisioned by Friston et al., if that outcome is what
CogNarr users desire. 

Lastly, although an explicit purpose of CogNarr is to facilitate cognition
in the large-group setting, one can also view CogNarr as a type of
Narrative Information Management (NIM) system, as described by \textcite{cordesNarrativeInformation2021}.
The function of a NIM is to manage a collection of narratives to support:
(a) identification of information gaps; (b) situational awareness;
(c) storage and recall of descriptive and explanatory information;
(d) compression of information, to reduce cognitive load (e.g., through
visualization and structuring); (e) case management; and (f) intelligence
synthesis. As such, the use of CogNarr narratives might extend beyond
group cognition per se. 

\section{Cognition}

Active inference, a Bayesian description of cognition and self-organization,
plays a central inspirational as well as technical role in the CogNarr
project. This section provides a brief overview of active inference
to give the reader a feel for how cognition is conceptualized in the
CogNarr ecosystem, including the purpose of cognition, how learning
occurs, why action and prediction are viewed as part of cognition,
and how action and prediction relate to beliefs. Communicating beliefs
and predictions is a core focus of the CogNarr ecosystem. This section
also briefly describes the hallmarks of basal cognition and the components
of group cognitive architectures, alerting the reader to question
how each hallmark is addressed in the CogNarr ecosystem and how functional
rather than dysfunctional cognition is favored.

\subsection{Active Inference }

\begin{figure}
\begin{centering}
\includegraphics[width=0.72\columnwidth]{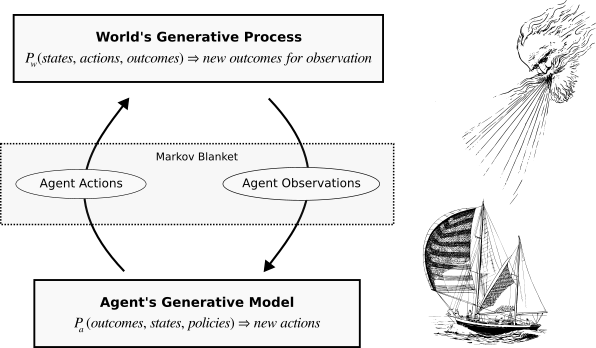}
\par\end{centering}
\caption{Active inference.\protect\label{fig:active-inference}}

\end{figure}

Active inference is a normative principle underwriting cognition in
biological and artificial intelligent agents \parencite{dacostaActiveInference2020,fristonActiveInference2017}.
As a brief sketch, consider two entities, an intelligent agent and
its world (see Figure \ref{fig:active-inference}). The agent, a small
entity relative to the massive world, cannot know everything about
the world. In particular, the agent cannot know the world's (hidden)
states and its generative process, $P_{w}$ in the figure. The world's
generative process is a joint probability function of states, agent
actions, and outcomes. The agent can only know its own actions and
what it perceives with its senses, which are noisy and partial observations
of the world's true states. The agent can, however, use its generative
model, $P_{a}$ in the figure, to estimate states of the world and
predict how the world might respond to an action or series of actions
(a \textit{policy} for action) that the agent entertains. The agent
uses its generative model to select new actions that it believes will
most likely achieve its aims. 

States of the agent and world only influence each other via observations
and actions. The agent and its world are said to be statistically
separated by a Markov blanket \parencite{kirchhoffMarkovBlankets2018}
comprised of observations and actions, as shown in Figure \ref{fig:active-inference}.
The state of the world is conditionally independent of the agent's
internal state, given agent actions, and the internal state of the
agent is conditionally independent of the world's state, given agent
observations. The situation in Figure \ref{fig:active-inference}
can be modeled, for example, by using a partially observed Markov
decision process (POMDP) \parencite{smithStepbystepTutorial2022,fristonFederatedInference2024a}.

Agent cognition can be described as an idealized four-part cyclic
process of \textit{predict $\rightarrow$ act $\rightarrow$ sense
$\rightarrow$ learn}, simplified here to be sequential and non-hierarchical.
Notably, actions are viewed as a component of cognition. Some actions
serve to gather information (thereby providing epistemic gain) while
others are goal oriented (thereby directly affecting conditions).
Given epistemic and goal-oriented motivations, an agent acts to reduce
its uncertainty about achieving and maintaining those preferred conditions
that (it expects will) bestow it with continued existence, wellbeing,
and homeostasis. Said differently, it acts in order to \textit{thrive},
where the term encompasses notions of wellbeing, resilience, and sustainability
discussed by \textcite{albarracinSustainabilityActive2024}. In short,
an agent favors choices that ensure the greatest resolution of uncertainty,
under the constraint that preferred outcomes are realized \parencite{fristonDesigningEcosystems2022}.

Prediction (anticipation) can result from a logical assessment of
a situation. For a human at least, it can also result from intuition,
the capacity to recognize and assess patterns without engaging logical
faculties or even conscious thought, or perhaps operating at the fringe
of conscious thought \parencite{priceIntuitiveDecisions2008,evansIntuitionReasoning2010}.
As such, an agent's generative model regarding some situation can
be more or less formal, so to speak. Regardless, if an agent's generative
model is faulty, the risk of making inaccurate predictions and/or
poor policy choices tends to increase. \textit{Learning} is the agent's
process of updating and improving its generative model to reduce predictive
errors and uncertainty. Ultimately, the \textit{predict $\rightarrow$
act $\rightarrow$ sense $\rightarrow$ learn} cycle is an optimization
process, where the objective function being minimized is expected
variational free energy.\footnote{Expected variational free energy represents the divergence between
predicted and preferred outcomes, minus the uncertainty (i.e., entropy)
expected under predicted states \parencite{fristonActiveInference2016}.
As such, free energy is minimized by increasing the accuracy predictions,
as well as by obtaining outcomes that align with preferred conditions
with certainty. Free energy minimization occurs at distinct temporal
scales that can be interpreted as: \textit{inference}, fast, short
term; \textit{model parameter learning}, medium term; and \textit{model
structure selection}, slow, long term \parencite{fristonFederatedInference2024a}.} Because free energy is bounded above by (Bayesian) surprise, it can
be minimized by minimizing surprise. Surprises occur when outcomes
do not agree with predictions and when uncertainty exists about whether
preferred states are or will be achieved. 

The above description of active inference applies to agents that are
individuals, such as an individual robot or human, as well as to agents
that are collectives, such as a group of humans. \textcite{guenin--carlutThinkingState2021},
for example, uses the active inference framework to examine City-States,
which he views as intelligent biological entities. 

\subsection{Functional/Dysfunctional Cognition and Cognitive Architectures }

Agents can and do take actions that are antithetical to thriving.
Some such actions are mistakes or miscalculations that an agent might
learn from. Others can form a pattern of behavior that poorly serves
or even harms the agent. Thus we can reasonably speak of functional
versus dysfunctional cognition, relative to the normative principle
of minimizing both predictive error and the uncertainty of achieving
and maintaining conditions that support thriving. Dysfunctional cognition
can arise from inadequacies related to any or all of the thirteen
hallmarks of basal cognition identified by \textcite{lyonCognitiveCell2015,lyonReframingCognition2021},
summarized below.
\begin{enumerate}
\item Valence (attraction, repulsion, neutrality) 
\item Sensing
\item Discrimination (ability to identify opportunities and challenges) 
\item Memory storage and recall 
\item Learning 
\item Problem solving (behavior selection, adaptability, abstract thinking) 
\item Communication (with others of the same or different kind) 
\item Motivation (teleonomic striving, implicit and explicit goals) 
\item Anticipation (prediction, forecasting, expectancy) 
\item Attention (oriented response, focus) 
\item Self-identity (distinguishing self from non-self) 
\item Normativity (error detection, behavior correction, value assignment) 
\item Intention (directedness toward an object, belief and desire)
\end{enumerate}
If functional cognition depends on the quality of hallmark processes,
it stands to reason that it also depends on the quality of the cognitive
architecture that prescribes and supports these processes. A cognitive
architecture is the set of components by which and through which cognition
occurs. Agent memory, for example, requires a storage mechanism, and
that storage mechanisms is part of an agent's cognitive architecture.
For a person, the cognitive architecture consists largely of a body-mind.
But tools of various kinds (books and computers, for example) or elements
in the environment (paths or arrangements of tools, for example) can
also be considered as part of a cognitive architecture \parencite{constantExtendedActive2022,whiteExtendingMinds2024}. 

For a group of humans, the cognitive architecture includes its members
(each with their own personal cognitive architectures), as well as
a set of norms, rules, mechanisms, institutions, tools, systems, and/or
procedures that facilitate or guide cognition \parencite{boikScienceDrivenSocietal2020a}.
And it includes the set of sensors and analytic processes that a group
implements to aid cognition. It follows that the CogNarr ecosystem
can be viewed as part of the cognitive architecture of the person
or group that is using it. 

In the group setting, dysfunctional cognition could arise from poor
or impoverished designs of the cognitive architecture, or from an
architecture that is poorly maintained or implemented. For example,
a dysfunctional architecture might motivate individuals to make hasty
or selfish decisions, provide insufficient information with which
to make wise decisions, tend to distract or scare individuals away
from important topics, offer poor or limited options during decision
making, coerce individuals toward certain decisions, or disempower
individuals such that their input does not matter, is limited to be
superficial, is difficult to convey, or is not requested. A cognitive
architecture might also be dysfunctional due to a low capacity to
remember past predictions, thus hampering a group's ability to assess
errors between predictions and outcomes. Or it might fail to adequately
sense, pay attention to, or process the information that is most important
for group wellbeing. The point is, a group cognitive architecture
has many potential modes of dysfunction, each related to one or more
hallmarks of cognition.

That being said, the CogNarr ecosystem is designed to facilitate functional
cognition, at scale. Given that the ecosystem can be viewed as part
of the cognitive architecture of the person or group using it, its
details matter. How does it address issues of attention, for example,
or of information recall? How does it achieve transparency and facilitate
distributed decision making? How does it incorporate predictions and
account for predictive errors? How does it communicate information?
The last question is tackled first.

\section{Story Graphs}

From an active inference perspective, group cognition rests on the
communication of (potentially dynamic and evolving) member beliefs,
and consensus building with respect to those beliefs. As described
by \textcite{albarracinSharedProtentions2024} for a generic group
in the normative setting, ``group members can be seen as actively
and implicitly aligning their beliefs and expectations through dialogue
and interactions, thereby enhancing their ability to predict each
other's actions and intentions, and thereby coming to perceive and
act in the world in similar ways.'' We can refer to a person's beliefs
and expectations as a \textit{belief model}, and equate it with $P_{a}$
in Figure \ref{fig:active-inference}. As such, a belief model is
an agent's internal model of how the world works, relative to some
situation. Two central tasks of CogNarr are to record a user's belief
model---externalize it, make it explicit and transparent---and then
to share it with others in order to facilitate group cognition. 

Most humans do not conceive of their own beliefs in terms of models,
however. Rather, humans tend to experience their beliefs and make
sense of the world largely through narratives \parencite{biettiStorytellingAdaptive2019,turnerMultifacetedSensemaking2023,fantirovettaDualRole2023,cordesNarrativeInformation2021}.
These can be internal narratives that a person constructs, adjusts,
and recites to himself or herself, or social narratives that are shared
within a group. In the active inference context, \textcite{albarracinVariationalApproach2021}
consider \textit{social scripts}, which are widely-supported prescriptions
about how one is to behave in various social settings, or what is
important in those settings. \textcite{bouizegareneNarrativeActive2020}
consider shared narratives conceived of more broadly. Social scripts
and shared narratives help humans to generate more accurate predictions
about the world and to coordinate social behavior. 

CogNarr builds on the human propensity to use narratives in sense-making
and, more generally, in cognition, by transforming internal narratives
of a particular type---those that reflect or resemble belief models---into
explicit, external belief models, which are then communicated. This
type of narrative is an account of events that conveys, for example,
a user's beliefs about what happened, why it happened, what it means,
what might happen in the future, what the user wishes would happen,
and what action should be taken, if any. 

Given the role of narratives in sense-making, CogNarr refers to the
externalized belief models as \textit{narratives}, or more simply,
as \textit{stories}.\footnote{The word \textit{story} is meant to imply an honest account of a user's
experience, beliefs, and expectations regarding some situation. It
is not meant to imply a fictitious, fanciful, insincere, or exaggerated
account. } Users might be encouraged to ``share your story,'' for example,
which seems more appealing than ``share your belief model.'' But
both phrases amount to the same thing. One might expect that stories
would be conveyed through written text, as is common outside of CogNarr.
But written text is unsuitable for the task at hand. CogNarr is designed
to facilitate cognition in the large-group setting, and cognition
at scale requires that stories exist in a format that is amenable
to automated computational assessment. No user is going to read, say,
10,000 written stories in order to digest and make sense of their
contents. Further, computers are not proficient at assessing the meaning
of written text, especially when that text is lengthy and reasonably
complex, and where its meaning must be understood at a deep level
(for example, where the logical implications of passages must be understood).
Thus, a story in CogNarr is not represented as text but rather as
a special type of computational graph, called a \textit{story graph},
which can be understood by both humans and computers. Story graphs
are the core innovation on which the CogNarr ecosystem is based. CogNarr
uses story graphs to curate, understand, and reason with high-quality
information about human experience, beliefs, and expectations.

Creating a story graph takes some effort on the part of a user, although
CogNarr is designed to make that process as easy and efficient as
possible. Further, creating a story graph is a foreign task that must
be learned, compared to creating a written document, which most people
are familiar with. So, why would someone want to spend time and energy
creating a story graph? Because doing so allows and facilitates group
cognition at scale. If one wants their potentially lengthy, complex,
or nuanced input to be heard and understood in the large-group setting,
then he or she might be willing to create their input in a format
that serves that purpose. 

The remainder of this section describes the purpose and components
of stories shared as story graphs, introduces some computational models
that could be used to conduct inference, describes how story graphs
are constructed and the tools used for construction, and describes
the process of editing story graphs, which can occur in rounds. Supporting
these processes is a backend computational system, which hereafter
is referred to as the \textit{system}. 

\subsection{Rich Stories }

The CogNarr ecosystem facilitates individual and group cognition by
helping a person to tell a potentially rich (e.g., nuanced, complex,
involved) story that conveys his or her experience, beliefs, and expectations
about a situation. A story can communicate a user's:
\begin{itemize}
\item Experience, understandings, views, questions, and concerns regarding
a situation, including explanations about what a situation means and
how and why it came to be. 
\item Predictions about about how a situation might unfold in the future,
given a series of actions or inactions, and predictions about what
might be learned through action. 
\item Beliefs about the understandings, intentions, motivations, and desires
of pertinent actors. 
\item Uncertainties about beliefs. 
\item Proposed solutions to a political, economic, environmental, technical
or other kind of problem, including proposed strategies to address
and/or avoid a problem.
\end{itemize}
Stories can be short, say, the equivalent of a paragraph in length,
or much longer, say, the equivalent of a book chapter. Let us assume
that an average story is the equivalent of one to several pages of
text in length. 

Hereafter, the term \textit{situation} refers to the situations, events,
circumstances, and/or problems that form the topic of a story. The
user is sometimes called a \textit{participant}, or \textit{group
member}. He or she might be called a \textit{storyteller} when creating
a story or a \textit{reader} when viewing one. 

CogNarr apps can be designed to support a wide range of use cases
in a wide range of domains for a wide range of groups. These include
use cases for individuals, apart from a group, and for small groups.
The focus of this paper, however, is on use cases that involve deliberation,
strategizing, collaborative problem solving, and/or decision making
in the large-group setting. These are among the most challenging of
use cases, and addressing them is a core purpose of the CogNarr project. 

The act of sharing stories and interacting about stories helps a group
to better understand the beliefs, concerns, desires, and expectations
of its members. Interactions can take several forms. For example,
group members and the system can pose questions or offer feedback
to a storyteller. A storyteller can edit his or her story as desired,
and can cite, comment on, use ideas from, or include portions of the
stories of other members. Likewise, a storyteller can create a story
from the synthesis of other stories. The blending of stories represents,
in the ideal, the blending of beliefs (the recognition and spread
of shared beliefs) within a group or subgroup. Depending upon group
rules, members can work in teams to create a story. 

\subsection{Story Components and Purpose }

In the CogNarr ecosystem, a completed story exists simultaneously
as:
\begin{enumerate}
\item A collection of passages. A \textit{passage} is written text, perhaps
one or a few sentences in length, conveying a set of closely related
thoughts. A passage also includes associated metadata and artifacts.
The collection of passages is considered the \textit{source of record},
from which all else is derived. 
\item A meaning representation of each passage, called a \textit{story graph
fragment.}
\item A meaning representation of the integrated set of passages, called
a \textit{story graph.}
\item A set of translations of the story graph, in whole or in part, into
natural language, alternative meaning representations, and/or computational
programs, called \textit{models}. 
\end{enumerate}
The purpose of the story graph is to: (a) represent a narrative in
a format that is less ambiguous than natural language; (b) serve as
a format that is readable and understandable by both humans and computers;
(c) provide a structure on which certain kinds of inference or analysis
can be applied (e.g., graph matching \parencite{faroukMeasuringSentences2020});
and (d) serve as an interlingua---an abstract natural-language-independent
representation suitable for translation into: (i) multiple natural
languages; (ii) one or more models, or arguments of models; and (iii)
alternative meaning representations that support specific models. 

Numerous meaning representations have been developed over the years
to support natural language processing and inference. Most of these
are graph-based. The reason for graphs is that they excel at representing
structured information, and both humans and computers are adept at
understanding information in graph form. CogNarr adopts a graph-based
meaning representation for the same reasons. Hereafter, for convenience,
the terms \textit{story graph} and \textit{meaning representation}
are sometimes used interchangeably. That is, a story in story graph
form is an instantiation of the CogNarr meaning representation. The
design of that representation is the same for story graphs and story
graph fragments, except that fragments might employ a more limited
set of features. 

\begin{figure}
\begin{centering}
\includegraphics[width=0.65\columnwidth]{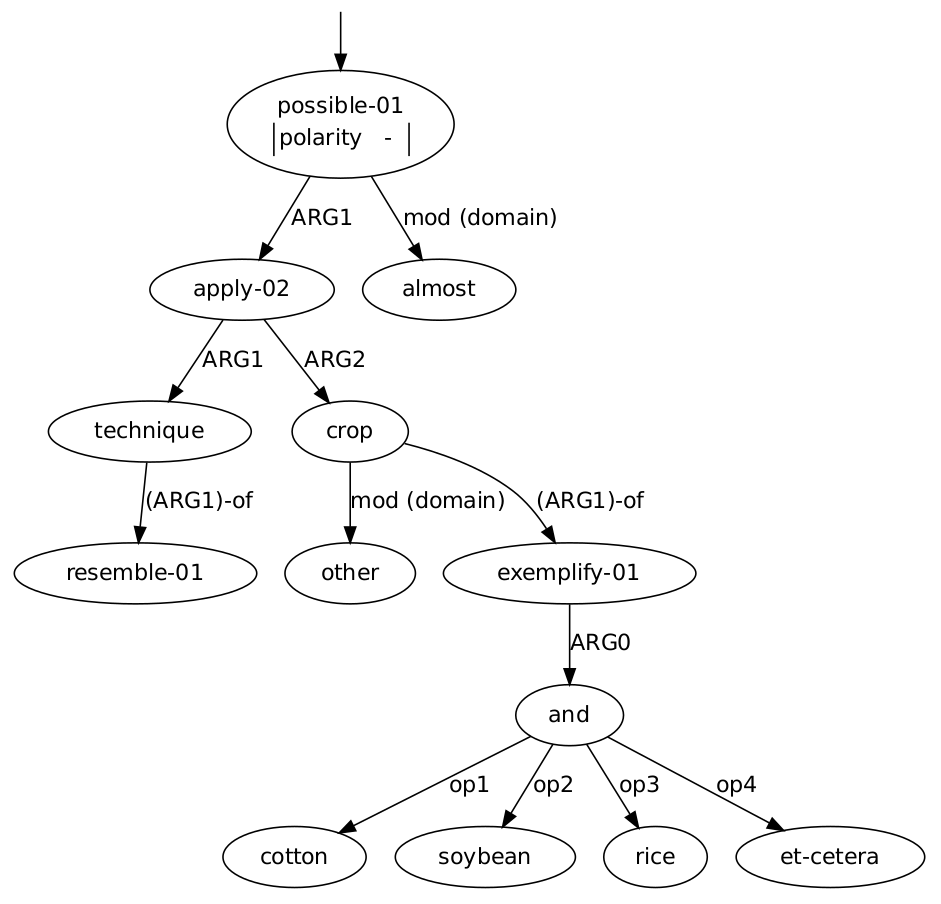}
\par\end{centering}
\caption{Abstract Meaning Representation for the sentence \textquotedblleft A
similar technique is almost impossible to apply to other crops, such
as cotton, soybeans and rice.\textquotedblright{} \protect\label{fig:AMR}}
\end{figure}

To give a flavor of existing meaning representations, the Abstract
Meaning Representation (AMR) \parencite{banarescuAbstractMeaning2013}
for an example sentence is reprinted from \textcite{oepenMRP20202020}
in Figure \ref{fig:AMR}. While AMR is a popular meaning representation,
CogNarr uses a custom representation that is designed to meet its
needs. Those needs and the CogNarr representation are discussed further
in \textcite{boikCogNarrEcosystem2024b}. 

\subsection{Models}

A story graph can be translated into one or more models. The purpose
of a model is to conduct inference or analysis, compute with uncertainties,
and/or generate predictions. \textit{Generate predictions} means to
instantiate a storyteller's beliefs regarding predictions. As an example,
if a storyteller conveys a belief of 100 percent certainty that the
next 30 days in the story world will be sunny, then the predictions
generated by a model will be a series of 30 forecasts, all of which
are for sun. If a storyteller believes that for each of the next 30
days there is a 50 percent chance of rain, then the predictions will
be a series of forecasts, about half of which are for rain. If the
model were executed a very large number of times, almost exactly half
of all predictions would be for rain. 

Two characteristic models are:
\begin{itemize}
\item A logic model that interprets formulas of classical or non-classical
logic. Classical logic, extensions of classical logic, and non-classical
logics include first-order logic, temporal logic, natural logic, probabilistic
logic, and annotated logic \parencite{bramerLogicProgramming2013,mossImplementationsNatural2018,abzianidzeLearningAbduction2020,ludlowLanguageForm2022,andreasenNaturalLogic2020,deraedtProbLogProbabilistic2007,liuReasoningBeliefs2021,adityaPyReasonSoftware2023,icardSimpleLogic2023,belleLogicProbability2020,decarvalhojuniorComprehensiveReview2024,abeThreeDecades2019}. 
\item A probabilistic model that is written in a universal probabilistic
programming language (PPL), such as the Julia package Gen \parencite{cusumano-townerGenGeneralpurpose2019}.\footnote{The author admits a bias toward the Julia programming language, due
in part to its speed, syntax, built-in parallelism, expressive type
system, and capacity for auto-differentiation. The software examples
provided in this document are almost all Julia packages. Much of their
functionality, but not necessarily all, is also available in other
programming languages, such as Python.} 
\end{itemize}
Other kinds of models are possible, some of which are mentioned next.
But for ease of exposition, we focus in this paper on translation
of a story graph, in whole or in part, into the two characteristic
ones mentioned, leaving the type of logic model generic. Further,
we assume, hereafter, that the translation of a story graph into a
model or other object is always \textit{in whole or in part}, as appropriate
to the situation. 

For practical reasons, only a small set of models would be included
in the initial CogNarr version. In a mature version of CogNarr, a
story graph could be translated into one or more of many possible
models, each serving a particular purpose. For example, a graph-matching
model could be used to assess the similarities between two or more
story graphs. A distributional model could be used to answer a query
about story meaning. The set of models constructed for a particular
story would depend in part on story characteristics, on the nature
of queries directed at the story, and on computational goals and resources. 

Continuing with possibilities, a model could be constructed in a specialized
PPL, rather than a universal one like Gen. For example, a story that
resembles a hidden Markov process, in whole or in part, might use
a PPL that is efficient for such cases. One such PPL is the Julia
package RxInfer \parencite{bagaevRxInferJulia2023}, which uses message
passing on factor graphs for probabilistic inference. In contrast,
a story that resembles a dynamical systems model might be translated
into a model from the ecosystem of Julia packages aimed at scientific
machine learning (SciML).\footnote{\url{https://docs.sciml.ai} }
That ecosystem supports ordinary and partial differential equation,
stochastic differential equation, neural-network, Bayesian network,
and jump process models, among others. 

As an alternative, a model could be constructed using the AlgebraicJulia
ecosystem,\footnote{\url{https://www.algebraicjulia.org}} which approaches
scientific modeling from an applied category theory perspective. That
ecosystem supports Petri-net, dynamical systems, and stock-flow consistent
models \parencite{baezCompositionalModeling2023,meadowsHierarchicalUpstreamDownstream2023}.
Applied category theory, along with (closely related) rich type theories,
is of special interest in this project and is discussed further in
\textcite{boikCogNarrEcosystem2024b}. 

Other possibilities include:
\begin{itemize}
\item Models in neuro-symbolic logic \parencite{chaudhuriNeurosymbolicProgramming2021,fengNeurosymbolicNatural2022}. 
\item Pre-trained language models (PTLMs) \parencite{huSurveyKnowledge2023,bommasaniHolisticEvaluation2023,myersFoundationLarge2023}.\footnote{In this paper, large language models, such as GPT-4, are referred
to as pre-trained language models (PTLMs). See \textcite{wangPreTrainedLanguage2023}
for a review.}
\item Bayesian and non-Bayesian statistical/distributional models \parencite{kalouliHyNLIHybrid2021,souzaHybridApproach2022,bernardyCompositionalBayesian2018,bernardyBayesianInference2023}. 
\item Models in other PPLs or that use other probabilistic approaches \parencite{brafmanProbabilisticPrograms2023,wongWordModels2023,dohanLanguageModel2022}. 
\end{itemize}
Finally, active inference models could be used \parencite{championDeconstructingDeep2023,ueltzhofferDeepActive2018,fountasDeepActive2020,devriesDesignSynthetic2023}.
For example, \textcite{albarracinSharedProtentions2024} suggest combining
a category-theoretic approach and active inference to model group
social actions based on shared goals. Further, because active inference
models can learn on the fly from observations, they might be employed
to initiate and orchestrate system interactions with storytellers.
One purpose of such interactions would be to help uses construct quality
(e.g., unambiguous, logically sound, on-topic) stories. 

Some models might require translation of a story graph into an alternative
meaning representation, for use as input. As an example, \textcite{kalouliGKRBridging2019}
combine structural and distributional features into a hybrid meaning
representation, which is then used as input to custom inference methods.
The final representation consists of five subgraphs layered on top
of a central conceptual (predicate-argument) graph. 

\subsection{Story Graph Views}

The CogNarr meaning representation is likely to be more full-featured
and sophisticated than AMR (Figure \ref{fig:AMR}) or other common
meaning representations. Thus, a long and complex story is likely
to be represented as a large and complex story graph. A typical CogNarr
user is not likely to be familiar with natural language inference
or the details of meaning representations, and so might have limited
use of viewing a raw meaning representation, especially for long stories.
On the other hand, a user might very well want to verify that his
or her story is being understood by the system as expected. For example,
suppose a storyteller enters the sentence ``Someone shot the friend
of the actress who was on the balcony.'' Perhaps the storyteller
knows what happened, but the system might be confused as to who was
on the balcony, the friend, the actress, or both? The system might
interpret the sentence in a way that the storyteller does not intend. 

If the storyteller suspects that there is more than one interpretation
of a sentence, or more than one way to combine a fragment into a story
graph, or has some other concern about how the story graph is being
constructed, then he or she might want to verify that the system understands
the story as intended. There are several ways that verification can
be done. One is to view a potentially simplified version of a story
graph fragment or story graph. Views can be tailored to meet the immediate
needs of the user. They can be simple, containing superficial, informal,
summarized, and/or partial information, or they can contain detailed
information. A fully detailed view is the underlying meaning representation
itself. 

\subsection{Uncertainty }

Assuming that a story is translated into a probabilistic model, the
model will be executed once every time a user instructs CogNarr to
generate an example outcome (unless a saved example is requested or
provided by default). In contrast, the model would be executed multiple
times, perhaps very many times, if a user instructs CogNarr to generate
a distribution over possible outcomes. Using the example already given
for a 50 percent chance of rain, if a user asks CogNarr how many of
the next 30 days are likely to be rainy, according to the story, CogNarr
would execute the model a large number of times and return the answer
15, if rounded to a whole number. Obviously, for simple problems like
this that have an analytic answer, CogNarr might not need to execute
the probabilistic model at all to produce an answer. It could execute
a more efficient model. As another example, suppose that a story has
a character, Alice, who usually rides her bike to work, but sometimes
drives her car. If a user instructed CogNarr to generate a set of
possible outcomes, in some outcomes (i.e., in some futures that the
storyteller conceives of), Alice will ride her bike to work. In others,
she will drive her car. 

If a story has certain outcomes, such as the sunny day example previously
given, predictions will be static. Each execution of the model produces
the same outcome. Such a story is deterministic. The same is true
for a story that does not include any uncertainty information, inferred,
implied, or explicit. Unless otherwise instructed, the system would
understand the lack of certainty information to mean that the storyteller
has complete certainty about outcomes. 

As in the examples with Alice and weather forecasting, a probabilistic
model might be very simple. In an uncomplicated short story containing
only one story-line branch, the model might represent not much more
than flips of a (fair or unfair) coin. If heads, Alice rides her bike
to work. If tails, she drives her car. More complicated stories would
involve more complex probabilistic models. 

For a given story, the storyteller, the readers, and the system can
all experience (or calculate) different levels of uncertainty. For
example, suppose a story has characters Alice and Bob and that the
storyteller knows Alice is Bob's friend, not his sister. But the storyteller
does not make the relationship clear in the story. Then a reader,
Carol, might query the story and ask about the relationship. The system
might respond, ``There is a 30 percent chance that Alice is Bob's
sister.'', which might be communicated in sentence or graph form.\footnote{Carol might also ask the system to explain how it came to this conclusion.}
Meanwhile, another reader, Dave, who does not issue a query, might
infer a different level of uncertainty, or might not even consider
that Alice could be Bob's sister. Carol and Dave, or the system, could
also directly ask the storyteller about the relationship. In this
case, if the storyteller replies, the answer should remove all uncertainty
about the relationship.

\section{Story Construction and editing}

\subsection{Story Construction}

\begin{figure}
\begin{centering}
\includegraphics[width=0.65\columnwidth]{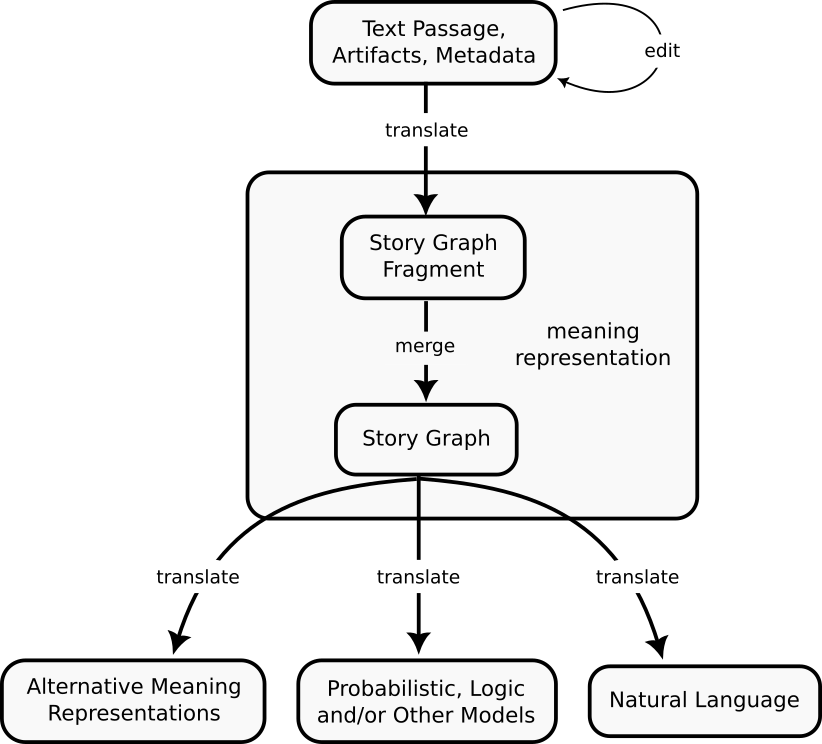}
\par\end{centering}
\caption{Story creation.\protect\label{fig:story-create}}
\end{figure}

A story is constructed one passage at a time, as illustrated in Figure
\ref{fig:story-create}. Each passage is translated into a story graph
fragment, and the fragment is merged into the growing story graph.
The story graph, in turn, can be translated into code for one or more
models, into alternative meaning representations, or into various
natural languages. For a story graph translated into model code, one
can think of the story graph as providing a semantic description of
code, perhaps even detailed elements of code, like algebraic equations.
If a logic model is employed, a story graph can be understood, roughly,
as a set of formulas within a logical-form language.\footnote{A logical form language is an encoding of the semantic content of
text that can be mapped to a logic-based representation. Thus, a formula
in a logical form language is an intermediary between raw text, which
has shallow semantic encoding, and a logical form, which has an expressive
semantic structure.}

When creating a story graph fragment, a storyteller can input sentences
as well as metadata and artifacts. In initial CogNarr versions, artifacts
would likely be limited to tables and charts. Sketch, import, and
other tools assist in their creation. The information in artifacts
is not just for human readers; it is integrated into the models and
so also into the meaning and predictions of a story. Over time, more
advanced versions of CogNarr could include images or other objects
as artifacts. 

Metadata are instructions or comments that help the system to understand
or process the meaning of input text. Metadata could indicate:
\begin{itemize}
\item That a passage has some rhetorical relation to another passage, such
as being an explanation or providing contrast. 
\item That a passage describes code, an algebraic function, or domain information.
\item Where or how a passage is to fit into the larger story, including
syuzhet instructions.\footnote{To the degree practical, a story graph is organized in canonical form
to facilitate computation. One way to do this, for some stories, is
to organize a story graph by fabula, a story's events ordered or indexed
chronologically. The storyteller's preferred presentation/telling
of the story, the syuzhet, might differ. Potentially, different layers
of a story graph could reflect fabula and syuzhet. }
\item That a passage is intended only for a specific model.
\item That the ambiguity of a sentence should be resolved in a particular
way.
\end{itemize}
To resolve ambiguity, metadata can include markup tags or symbols
within a sentence that indicate the intended semantics. In some cases,
markup could be as simple as use of brackets to indicate scope (e.g.,
``Someone shot the friend of {[}the actress who was on the balcony{]}''). 

As mentioned, the set of passages becomes the source of record from
which all else derives. Consider the simple case where a passage is
only text, as entered into a text box. In this and more complicated
cases, it is useful to have a single source of record. If multiple
sources exist (e.g., if the user can edit passages, story graph fragments,
and the story graph), then information in the various sources might
conflict. At the very least, the new information in each edit would
need to automatically propagate across sources. With passages as the
single source of record, edits propagate in one direction only (i.e.,
downward in Figure \ref{fig:story-create}). Also, the user is always
editing his or her own input, rather than editing derivations that
the system makes as it propagates information from one source to another.\footnote{While this is true, the situation is a bit more complicated. Suppose
the system asks the user about the meaning of text, say, to resolve
anaphora. The user might respond with something like, ``\textit{He}
refers to Bob.'' That information also has to be part of the source
of record or else the story meaning could not be fully reconstructed
from the source. If the system only alters the story graph after receiving
the user's response, then there would be two sources of record, the
original passage and the system-altered story graph. To avoid this
problem, the system saves the question and the user's response as
part of the source of record, as metadata. The system or user might
need to remove or alter this metadata if the story changes later.
Alternatively, in this example, the storyteller could just change
the word \textit{he} to \textit{Bob} in the passage, and no metadata
would need to be saved. }

\subsection{Tools for Story Graph Construction}

Story graph construction is assisted by tools and guided by system
feedback. Passages are automatically translated into story graph fragments,
and when a translation is not possible, the system might ask questions
or offer suggestions. Current text-to-graph parsing methods \parencite{abzianidzeFirstShared2019}
employ neural and statistical methods \parencite{liuUnderstandingGenerating2021,vannoordExploringNeural2018,ozturelCrossLevelTyping2022,kalouliHyNLIHybrid2021},
commonsense resources such as knowledge graphs \parencite{ilievskiDimensionsCommonsense2021},
syntactic parsers \parencite{poelmanTransparentSemantic2022}, and/or
PTLMs \parencite{panUnifyingLarge2023,biCodeKGCCode2023,wangPreTrainedLanguageMeaning2023}.
Current graph-to-code methods also employ PTLMs \parencite{wangHypothesisSearch2023,tipirneniStructCoderStructureAware2023}. 

Merging a fragment into a story graph is automatic, guided as necessary
by user-supplied metadata and questions and suggestions offered by
the system. To verify that the system understands the intended meaning
of input, the storyteller can view the story graph, as mentioned,
can view predictions and outcomes, and/or can translate the story
into natural language. An advanced version of CogNarr might generate
animations or other multimedia as part of story playback. Current
graph-to-text generation methods employ PTLMs \parencite{ribeiroInvestigatingPretrained2021,ribeiroStructuralAdapters2021}
or other kinds of neural networks \parencite{liuTextGeneration2021}. 

\subsection{Why Story Graphs and Fragments?}

One might wonder why both story graph and model representations are
necessary. Why not convert natural language directly into probabilistic
or other kinds of models, if that is the goal? There are several reasons.
First, a model might represent only a portion of a story graph. Second,
humans are likely to grasp the meaning of a graph more easily and
faster than the meaning of code. Coding requires special skills, whereas
understanding a story graph view generally does not. Third, some computational
methods are more efficient or simpler on graphs, compared to code.
For example, scoring the similarity between two graphs is a common
operation. Methods like graph neural networks excel at processing
semantic or other information stored as graphs \parencite{wuGraphNeural2023}.
Fourth, a graph that contains semantic information about code can
improve accuracy on automated tasks such as code assessment (e.g.,
bug identification), summarization, and documentation \parencite{allamanisLearningRepresent2018,jiangTreeBERTTreeBased2021}.
Fifth, a story graph can serve as an interlingua, whereas achieving
this with code would be more challenging. 

Why are story graphs created fragment by fragment? By doing so, the
user breaks up the text-to-graph translation process into shorter,
more manageable steps. Furthermore, the step-by-step approach offers
opportunity for real-time system-user interaction to reduce story
ambiguity and correct errors. For example, after constructing a graph
fragment from text, the system might prompt the user, ``I am confused.
Is Alice Bob's sister, or his friend?''

\subsection{Editing in Rounds}

\begin{figure}
\begin{centering}
\includegraphics[width=0.4\columnwidth]{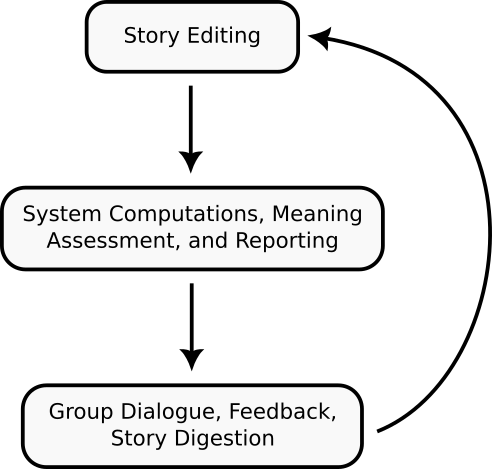}
\par\end{centering}
\caption{Story evolution in rounds of editing.\protect\label{fig:story-rounds}}
\end{figure}

In some cases, a group will want to edit stories over a series of
rounds, as illustrated in Figure \ref{fig:story-rounds}. In each
round:
\begin{enumerate}
\item Users edit their stories, with possible feedback from the system. 
\item The system processes all stories and reports back to the group. For
example, it reports on similarities and differences between stories. 
\item The group considers the meaning of stories and engages in dialogue
and story feedback. 
\end{enumerate}
As rounds progress, ideally the group comes to better understand the
perspectives of its members, and, in the decision-making case, hone
in on a well-supported, favored decision. That decision is itself
in the form of a story graph. In other words, the narratives of the
group ideally evolve toward one or a few central narratives that best
represent the shared understanding of the group (and potentially,
also any divergence of understanding). It is worth noting that a selected
decision could, in some cases, be in the form of a policy that outlines
a series of actions. That policy could include if-then statements. 

Reporting by the system in each round helps the group to better understand
the range of stories, the characteristics and qualities of stories,
and the evolution of stories during rounds. By pointing out commonalities
and differences among stories (for example, based on quality, content,
time frames, and predictions), clustering stories by similarities,
and by reporting other characteristics of the set of stories, the
system helps draw attention toward potentially salient features. Further
development of attention mechanisms in CogNarr is left for future
work. 

The three-step process illustrated in Figure \ref{fig:story-rounds}
makes group cognition explicit and transparent. Further, it generates
a history of narrative evolution that can be recalled and used for
learning and evaluation. The quality of the group cognitive process
itself could be evaluated. The history also serves as a record of
majority and minority opinions, which could be evaluated by the group
in hindsight. For example, evaluation could assess the accuracy of
predictions made by different subgroups. 

\subsection{Story Repository}

As a result of the editing, feedback, and assessment processes described
above, story ambiguity is minimized. Most finished stories would consist
of rich, high-quality, curated information about human experiences
and beliefs. Stories that are low quality could readily be identified.
If CogNarr is successful, over time the accumulated volume of high-quality
narratives would become massive. Such data would be valuable to academic
research (in fields such as sociology, governance, psychology, and
cognition). And it would be valuable to the system itself, which could
use it for continuous learning. As the data repository grows, so too
would the system\textquoteright s capacity to understand and analyze
both stories and strategies for action. Data privacy, data ownership,
and data control are critically important concerns. Addressing these
topics is left for future work. 

\section{Storytellers and Active Inference }

The preceding sections have introduced both active inference and stories.
Now we take a look at storyteller relationships in the context of
active inference. These relationships suggest ways that active inference
naturally occurs in the CogNarr setting, and as well suggest ways
that active inference models could be used to assist users.

Referring back to Figure \ref{fig:active-inference}, active inference
is a normative process that occurs between an agent, which we now
label \textit{Self}, and its world. The world can include other agents,
and we now label it \textit{Other}. Self is trying to understand Other
and Other is trying to understand Self. From an active inference perspective,
interactions between Self and Other that involve honest communication
can help agents develop common beliefs and common expectations about
what is likely to happen \parencite{vasilWorldUnto2020}. Several
studies have framed belief-sharing and cooperation between agents,
including cooperative communication, in an active inference context
\parencite{maistoInteractiveInference2023,albarracinEpistemicCommunities2022,poppelResonatingMinds2022,fristonFederatedInference2024a,albarracinSharedProtentions2024}.

\begin{figure}
\begin{centering}
\includegraphics[width=0.72\columnwidth]{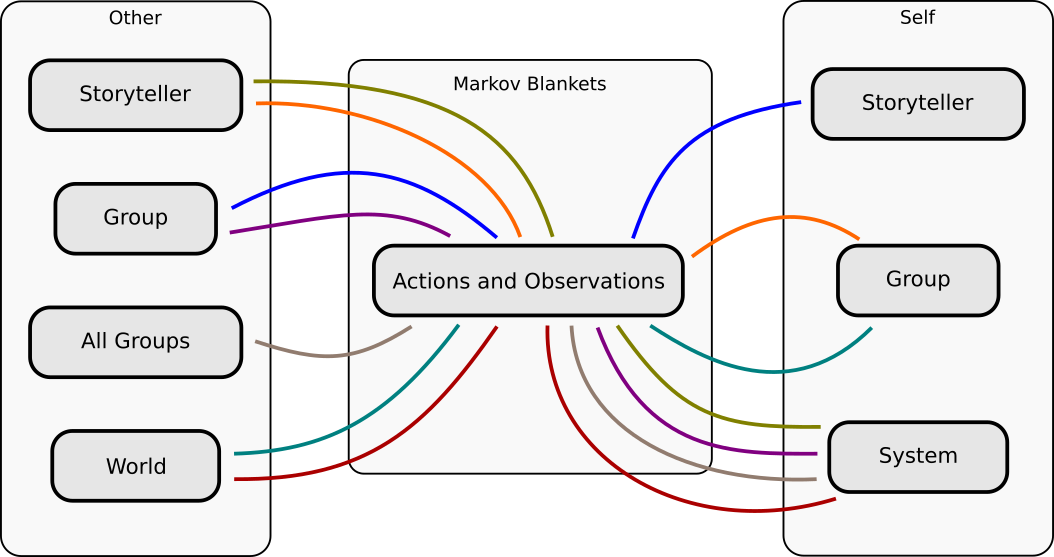}
\par\end{centering}
\caption{Active inference relationships.\protect\label{fig:actinf-relate}}
\end{figure}

In the CogNarr setting, multiple Self-Other relationships can be identified.
Some examples are illustrated in Figure \ref{fig:actinf-relate}.
A storyteller tries to understand a group. The system tries to understand
a storyteller, the group, and the set of all groups. A group tries
to understand a storyteller. And a group tries to understand the world,
in part though member narratives and, in the decision-making case,
in part by choosing actions, implementing them in the world, and then
sensing and evaluating outcomes. 

In all these cases, an agent is choosing actions based on beliefs
and predictions. Some actions are intended to gain information and
some are intended to alter conditions. Those human-human and human-world
interactions illustrated in Figure \ref{fig:actinf-relate} that are
informal, not part of a mathematical model, are examples of active
inference in the wild. Fox calls this \textit{implicit} active inference
\parencite{foxAccessingActive2021}.

Formal active inference can occur in CogNarr too. As mentioned, an
active inference model could be used to orchestrate system-initiated
interactions that are intended to assist and guide storytellers. In
this case, the system uses an active inference model to refine its
understanding of a story and storyteller, with the goal of improving
story quality. Among other things, the system tries to reduce its
uncertainty about events and characters in a story, and reduce any
discrepancies between the story content and rules or guidelines that
the storyteller is asked to follow. The system's preferred conditions
include high-quality stories that are appropriate for the situation
at hand. As an example, storytellers in a group might be asked to
stay on topic, and the system might occasionally remind a user of
this if he or she drifts too far away. 

To assist and guide users, the system should be capable of asking
intelligent questions about a story, the kind of questions that a
thoughtful and respectful human active-listener would ask. These include
questions about who the characters are, where the story is going,
what the storyteller intends, and what pieces of information are missing
that, if obtained, would most reduce story ambiguity or improve clarity.
To ask such questions, the system must be able to understand a story
in its current form and predict not only what might be gained if certain
new information were added or existing information altered, but also
how its prompts or questions might best achieve the aim of eliciting
a desired change in information. In order to learn, the system must
be able to assess any prediction error between expected and observed
user responses, after it takes some action. 

In later versions of CogNarr the role of active inference models could
be expanded. For example, in CogNarr versions that allow for game/role
playing and strategy development, the system might use an active inference
model to help a user develop or explore a policy for action, or help
a user to identify the goals of other players or characters. Planning
and goal identification are topics of active inference research \parencite{kaplanPlanningNavigation2018,matsumotoGoalDirectedPlanning2022,pezzatoActiveInference2023}. 

In the game/strategy setting, the system might play the role of one
or more characters in a story, some of whom are in a cooperative,
uncooperative, or antagonistic relationship with the story's protagonist.
Among other things, the system might predict a character's next move.
In an interesting twist on this theme, the system could potentially
play against itself, as if there were two systems playing, representing
two different protagonists. Here, story graphs might be used as a
language for (a type of) inter-system communication.

As another example for later CogNarr versions, the system might use
an active inference model to help steer a decision-making group toward
a genuinely well-supported decision. From the perspective of the system,
the \textit{act} step of the \textit{predict $\rightarrow$ act $\rightarrow$
sense $\rightarrow$ learn} cycle might represent a series of probes
and suggestions to the group. Of course, such a steering process would
need to be transparent. But the concept is not novel. Human facilitators
regularly perform similar tasks in the small-group, face-to-face setting. 

\section{Ecosystem Components }

In summary of what has been said so far, core components of the CogNarr
ecosystem include:
\begin{itemize}
\item Use-case-specific frontend apps that allow narratives to be constructed,
shared, viewed, discussed, and updated. A variety of tools assist
these processes. 
\item A backend system that allows for narrative storage and recall, as
well as networking and computational services. Computational services
include:
\begin{itemize}
\item Construction of story graph fragments from natural language sentences. 
\item Integration of fragments into a larger story graph. 
\item Translation of story graphs into logic, probabilistic, and/or other
models, and execution of those models to generate predictions and/or
perform inference and analysis. 
\item Translation of story graphs into natural languages. 
\item Computational assessment of narratives (e.g., for content, meaning,
quality, characteristics, and similarity) 
\item Processing queries directed to stories or storytellers regarding story
meaning, reasoning, or predictions. Both users and the system can
generate queries. 
\item Providing interactive guidance to storytellers to help them create
high-quality stories.
\end{itemize}
\end{itemize}
As mentioned, the CogNarr ecosystem is conceived to be an open-source
project. At the organizational level, it would employ some kind of
governance and administrative system. In addition to developing CogNarr
itself, the project might also provide services and outreach to increase
awareness, educate the public, and train individuals and groups to
effectively use CogNarr. And the project might support research that
furthers cognitive science and contributes to the effort of developing
tools to facilitate individual and group cognition. 

\section{CogNarr in Practice}

Having described stories and the process of story creation and editing,
we now turn to some practical aspects about how groups form, the function
of stories, how story quality is measured, and how cooperative user
behavior is encouraged. Example use cases are discussed to pique the
reader's imagination as to how CogNarr applications might be used
in the world. 

\subsection{Story Events }

Users create stories in response to and as part of a \textit{story
event}. There are at least two kinds of story events: \textit{self}
and \textit{group}. A self event is created by an individual and only
that individual participates. A person might create a self event to
strategize about a problem, to help clarify his or her thoughts about
a situation, or to create a permanent record of beliefs and information,
for later use. Self events might grow in popularity if and when CogNarr
versions allow for game/role playing and strategy development. 

A group event is one that is announced and hosted by a group.\footnote{\textit{Group} can refer to an ad hoc or standing group that hosts
events, or to the group of users that is engaged in a specific event.
The distinction should be clear from context.} For example, a community organization might host a group event to
generate potential solutions to a community problem. The group would
invite users to join the event, define rules for the event, describe
the purpose of the event, and provide other necessary information.
If the purpose of the event is decision making, then the group would
define the methods, rules, and procedures to be used. A variety of
stock methods are available for decision making including majority
rule, ranked choice voting, score voting, and dynamic representation,
as used in liquid democracy \parencite{valsangiacomoPoliticalRepresentation2021}.
It is worth noting that a group could, if desired, split a problem
into pieces that are more easily addressed, and then form subgroups
to address certain parts of a problem.

The above are some examples of the flexibility that CogNarr provides
for group organization, methods, procedures, approaches, and rules.
In addition, groups decide upon member anonymity---member identity
can be hidden or displayed to the group. Groups themselves, however,
must be transparent. For example, the organizational structure of
a group, its methods, procedures, and rules, and perhaps also its
funding and other selected characteristics must be transparent to
group members. 

Whether group or self, an event can have a brief duration (say, 24
hours or less), a prolonged duration (weeks or months), or be ongoing,
such as might occur for a situation that is evolving over time and
that needs ongoing evaluation or reporting. 

A group event can involve just a few people or potentially millions.
Further, communications can be many-to-many (e.g., collaborative decision
making), many-to-one (e.g., polls, crime reporting, and customer returns),
or one-to-many (e.g., product instructions or public announcements). 

\subsection{Story Function and Quality }

The function of a story is to convey a storyteller\textquoteright s
honest beliefs and expectations about a situation. The storyteller's
group might like to know how the beliefs and expectations contained
in a story align with objective reality. But objective reality can
be difficult, if not impossible, to define. So, the system does not
try to make this assessment, except in the case where the group provides
explicit instructions to the system as to how it should evaluate story
realism. Each group member can also assesses the realism of a story,
as desired, and a group can appoint committees or other bodies to
assist in evaluations. 

As an absurd example, suppose that a group event is focused on solving
a local transportation issue, say, traffic congestion. Bob submits
a story in which cars travel at warp speed. While Bob\textquoteright s
story might be creative, members of the group are not likely to value
Bob\textquoteright s contribution because they would view it as unrealistic.
An evaluation committee, if one exists, might agree with this assessment.
A reader might question Bob by asking, ``Do warp-speed cars exist
today?'' Another reader might add to their own story, referencing
Bob's, conveying a belief that warp-speed cars do not exist and so
are not a good solution to the current problem. Further, the system
might flag the story as containing futuristic components, if given
instructions on how to do so.

Likewise, group members are responsible for assessing the predictive
accuracy of a story, as time unfolds. Did a situation evolve as a
story predicted? Again, a group can appoint committees or other bodies
to assist in evaluations, and the system can perform automated assessment
if given explicit instructions on how to do so. The system also provides
aids for assessing predictions, including analytic tools and the saved
history of stories themselves. And of course, it provides communication
tools that can be used to disseminate any evaluations. It can also
maintain a history of such evaluations.

The focus on conveying a storyteller's honest beliefs about a situation,
rather than claims about objective reality, frees the user from having
to prove, in some sense, how the word works or what exists in the
world. By focusing on beliefs, a storyteller might be able to convey
his or her understandings merely by providing a guesstimate (a semi-quantitative,
best-guess) about conditions. For example, a storyteller might believe
that the economy will slow down (or improve) if taxes are raised on
the wealthy. The expected trend could be conveyed in a simple chart,
such as the kind scratched on the back of a napkin. CogNarr apps include
a sketch tool for this purpose. 

Ideally, a storyteller provides information in the right volume and
quality to make a story sufficiently clear, concise, consistent, logical,
and complete. These, in turn, help make the story useful to the storyteller
and group. Naturally, stories and beliefs that are based on sufficient
evidence and sound reasoning would tend to be more persuasive to group
members than those based on thin evidence or unsound reasoning. 

In some cases, the situation at hand will be one that experts have
studied intensively. Solid data and good mathematical models or other
kinds of information might exist to support or oppose a particular
belief. If so, a storyteller can reference existing information or
otherwise include that information within a story. A very sophisticated
story might directly incorporate a formal model of some worldly process.

While the system does not assess story realism (unless given instructions
on how to do so), or predict what will happen in reality if certain
actions are taken or not taken, it does assess story quality. This
assessment centers on how comprehensible, complete, internally consistent,
timely, and relevant a story is likely to be to a group, given the
event\textquoteright s purpose. This is termed the \textit{intrinsic
quality} of a story. One reason to assess intrinsic quality is to
provide a storyteller with feedback about his or her story, as it
evolves.

Regarding completeness, are there missing elements or gaps in reasoning,
motivation, or action that render the story difficult to understand?
Regarding consistency, does the story exhibit an internal logic? Are
characters sufficiently defined and reasonably consistent? Regarding
timeliness and relevance, does the story address the target situation?
Does it conform with any rules defined for the event? A group might
have other concerns regarding intrinsic story quality, and if so,
the group can instruct the system to consider these also when assessing
quality. 

\subsection{User Behavior, Rules, and Reputation }

The set of rules and policies adopted by CogNarr and by groups should
serve to support and encourage:
\begin{itemize}
\item An enjoyable, safe, and engaging user experience.
\item Healthy and honest communication.
\item The efficient creation of high-quality stories, containing high-quality
information.
\item Ultimately, highly functional group cognition and the empowerment
of groups to make a positive difference in the world. 
\end{itemize}
For example, a CogNarr policy could specify that a single human can
have only one CogNarr account at a time (or perhaps only one account
per app at a time), or a maximum number of accounts over a period
of time. Another policy could specify that the account holder must
be provably human (i.e., not a bot). Such rules might reduce the capacity
of an actor to unfairly manipulate group events.

\begin{table}
\begin{tabular}{|>{\raggedright}p{0.2\columnwidth}|>{\raggedright}p{0.78\columnwidth}|}
\hline 
{\small\textbf{Use Case}} & {\small\textbf{Narrative Snippet}}\tabularnewline
\hline 
\hline 
{\small A. Decision making} & {\small To address issue $X$, we must first address issue $Y$ that
caused $X$. Issue Y originated when \_\_\_, and has had the effect
of \_\_\_. If we don\textquoteright t solve $Y$, then event $A$
is likely to occur within \_\_\_ months and $X$ will be impossible
to solve. If $A$ happens, then \_\_\_ could be reduced, perhaps by
20 percent or more. A possible solution to $Y$ is to do \_\_\_ if
event $B$ occurs and \_\_\_ if event $C$ occurs.}\tabularnewline
\hline 
{\small B. Customer service} & {\small Your product $A$ arrived with part $C$ missing. I fixed it
with something I purchased at the hardware store. But the bigger problem
is that after using your product under condition $F$, a leak in part
$G$ caused part $H$ to slip. It\textquoteright s not that part $G$
was broken, but rather it was so thin it did not fit properly in slot
$J$. Was my part $G$ defective?}\tabularnewline
\hline 
{\small C. Emergency reports} & {\small I just witnessed an accident on the north side of intersection
\_\_\_. A large truck, it looks like a tanker of some kind, began
swerving. Perhaps it swerved because of the wet conditions. It shouldn't
have been going so fast. There have been many accidents at that corner.
The truck hit an electric pole and the wires started sparks in the
grass. Just to the north is a school and children are playing outdoors.
I\textquoteright m about half a block away, to the East, so all of
this is a bit hard to see.}\tabularnewline
\hline 
{\small D. Public polling} & {\small I have watched candidate $X$ for years. I thought her decisions
on issues $A$ and $B$ were terrible and I've been fuming about them
ever since. She should have known that her decisions would lead to
\_\_\_. As a candidate, I don\textquoteright t trust her on issue
$C$ because of her statements \_\_\_. Frankly, I don\textquoteright t
believe her on that issue, and I\textquoteright m not even sure she
is sincere. Besides, I think that issue $C$ should be handled by
actions $D$, which are inexpensive and easy to apply. I am unlikely
to vote for her.}\tabularnewline
\hline 
{\small E. Instructions} & {\small Product $A$ can be used for purposes $B$ and $C$. If used
for purpose $B$ and condition $D$ is present, then safety value
$E$ should be engaged whenever others are within distance of \_\_\_.
For purpose $C$, the fuel mixture should be \_\_\_, with no more
than 10 units of \_\_\_.}\tabularnewline
\hline 
{\small F. Personal strategizing} & {\small I'm confused as to why person $A$ sent item $B$ to person
$C$, and why it occurred before date $D$. It makes sense to me that
person $C$ might have needed item $B$ to finish a project, but I've
been told this is usually handled by person $E$. Perhaps person $A$
wanted person $C$ to think that \_\_\_. As a strategy, I will take
action $F$ to see if event $G$ occurs. That might tell me if \_\_\_
is true. If only I had taken \_\_\_, I would know what person $A$
is doing.}\tabularnewline
\hline 
{\small G. System description (analytic)} & {\small Rabbit population, $r$, is prey for a population of wolves,
$w$. Let parameters $\alpha,\beta,\gamma,\delta>0$. The change in
the rabbit population over time period $dt$ is $\frac{dr}{dt}=\alpha r-\beta rw$.
The change in the wolf population is $\frac{dw}{dt}=-\gamma w+\delta rw$.}\tabularnewline
\hline 
{\small H. System description (semi-quantitative)} & {\small Rabbit population, $r$, is prey for a population of wolves,
$w$. If $r$ increases, then $w$ will start to increase. But as
$w$ increases further, $r$ will start to decrease, which eventually
leads to a decrease in $w$. Populations $r$ and $w$ rise and fall
in cycles, as suggested by the attached rough sketch.}\tabularnewline
\hline 
\multicolumn{1}{>{\raggedright}p{0.2\columnwidth}}{} & \multicolumn{1}{>{\raggedright}p{0.78\columnwidth}}{}\tabularnewline
\end{tabular}

\caption{Potential use cases and narrative snippets.\protect\label{tab:use-cases}}

\end{table}

Defining and enforcing an optimal mix of rules and policies is a substantial
challenge. Almost certainly, some users will, at times, or consistently,
behave in a fashion that threatens to degrade group cognition. In
some cases, intentions will be benign. In other cases, intentions
will be to manipulate, exploit, mislead, or otherwise harm users or
disrupt group cognition. The rules and policies adopted by CogNarr
and groups should encourage cooperative behaviors and discourage uncooperative
and detrimental behaviors. Any such rules and polices should be fair
and transparent, and should provide opportunity for recourse if one
believes that a judgment about infringement has been made in error.
Addressing these challenges is left for future work, except for a
few thoughts, on reputation, which are mentioned below. 

Reputation can be a useful form of social currency in the CogNarr
ecosystem. But how should reputation be assessed, based on what information,
and by whom? And how can one's reputation be renewed or repaired?
These are challenging questions, and like other challenging questions
about rules and policies, they are left for future work. Note, however,
that members of a group will naturally tend to form opinions about
other members, and act accordingly. Indeed, it is generally wise for
a user to give more weight to a story that is created by a person
who has a reputation for making thoughtful, insightful contributions,
communicating honestly, listening to others, adhering to group and
CogNarr rules, predicting the future accurately, and treating others
with respect. Some groups are large, some groups are public, and in
some groups members are not anonymous. As such, one's reputation could
potentially be assessed across a very large population. 

Beyond user-driven reputation assessment, it might be useful for CogNarr
and/or groups to perform certain types of assessment. For example,
StackOverflow conducts a limited form of reputation assessment based
on adherence to community guidelines and engagement.\footnote{\url{https://stackoverflow.com/help/whats-reputation}}
More sophisticated approaches are certainly possible.

\subsection{Use Cases }

What sorts of stories can be created in CogNarr, and for what purposes?
To give a flavor of possible use cases, several examples with attendant
snippets from imagined user stories are provided in Table \ref{tab:use-cases}.
A dedicated CogNarr app could be developed for each use case, and
for specific domains. While the snippets in the table contain multiple
passages, actual stories would be written in the process depicted
in Figure \ref{fig:story-create}. In the snippets, blanks and variables
($A$, $B$, etc.) are used as placeholders. 

\begin{table}
\begin{centering}
\begin{tabular}{|>{\raggedright}p{0.25\columnwidth}|>{\raggedright}p{0.75\textwidth}|}
\hline 
{\small\textbf{Phenomena}} & {\small\textbf{Example Passage}}\tabularnewline
\hline 
\hline 
{\small Conditional and deontic modality} & {\small\textbf{Use Case E:}}{\small{} ``If used for $B$ and condition
$D$ is present, then safety value $E$ should be engaged''}\tabularnewline
\hline 
{\small Anaphora} & {\small\textbf{Use Case D:}}{\small{} ``She should have known that
her decisions would lead to \_\_\_'', with she and her referring
to candidate $X$}\tabularnewline
\hline 
{\small Numbers/arithmetic} & {\small\textbf{Use Case E:}}{\small{} ``no more than 10 units of \_\_\_''}\tabularnewline
\hline 
{\small Tense and coordination} & {\small\textbf{Use Case D:}}{\small{} ``I thought her decisions on
issues $A$ and $B$ were terrible''}\tabularnewline
\hline 
{\small Evidentiality} & {\small\textbf{Use Case F:}}{\small{} ``but I've been told that this
is usually handled by''}\tabularnewline
\hline 
{\small Negation} & {\small\textbf{Use Case B:}}{\small{} ``It\textquoteright s not that
part $G$ was broken''}\tabularnewline
\hline 
{\small Hyponymy} & {\small\textbf{Use Case C:}}{\small{} ``large truck, it looks like
a tanker'' A tanker is a kind of truck.}\tabularnewline
\hline 
{\small Polysemy} & {\small\textbf{Use Case D:}}{\small{} ``and I've been fuming about
them ever since'' Here, fuming suggests anger, not a smoking object.}\tabularnewline
\hline 
{\small Meronymy} & {\small\textbf{Use Case B:}}{\small{} ``a leak in part $G$ caused
part $H$ to slip\textquotedblright{} Here, $G$ and $H$ are components
of Product $A$.}\tabularnewline
\hline 
{\small Counterfactual reasoning} & {\small\textbf{Use Case F:}}{\small{} ``If only I had taken \_\_\_,
I would know what person $A$ is doing''}\tabularnewline
\hline 
{\small Spatial deixis} & {\small\textbf{Use Case C:}}{\small{} ``Just to the north is a school''
This entails that the school likely abuts the accident scene.}\tabularnewline
\hline 
{\small Parenthetical context} & {\small\textbf{Use Case C:}}{\small{} ``There have been many accidents
at that corner.''}\tabularnewline
\hline 
{\small Uncertainty} & {\small\textbf{Use Case C:}}{\small{} ``I\textquoteright m about half
a block away, to the East, so all of this is a bit hard to see.''}\tabularnewline
\hline 
{\small Differential equation} & {\small\textbf{Use Case G:}}{\small{} The entire passage describes a
differential equation.}\tabularnewline
\hline 
{\small Approximation} & {\small\textbf{Use Case H:}}{\small{} ``as suggested by the attached
rough sketch'' If the sketch contains numerical values, they are
to be interpreted as approximate.}\tabularnewline
\hline 
{\small Comparatives} & {\small\textbf{Use Case B:}}{\small{} ``it was so thin it did not fit
properly in slot $J$'' }{\small\textit{Thin}}{\small{} here is relative
to the width of slot J.}\tabularnewline
\hline 
{\small Common Sense} & {\small\textbf{Use Case D:}}{\small{} ``The truck hit an electric pole
and the wires started sparks in the grass.'' This entails that the
wires fell down from the pole and onto the grass.}\tabularnewline
\hline 
\multicolumn{1}{>{\raggedright}p{0.25\columnwidth}}{} & \multicolumn{1}{>{\raggedright}p{0.75\textwidth}}{}\tabularnewline
\end{tabular}
\par\end{centering}
\caption{Selected linguistic and knowledge-representation phenomena from use
case examples. \protect\label{tab:linguistics}}
\end{table}

The use case examples in Table \ref{tab:use-cases} are by no means
exhaustive. Other use cases might focus on education, for example.
Perhaps a group wants to learn about and explore a topic, for epistemic
value. Personal strategizing is listed as a use case in the table.
But perhaps a group wants to develop strategies to influence change
in the world. In some cases, that group's cognitive process might
involve numerous inner and outer rounds of \textit{predict $\rightarrow$
act $\rightarrow$ sense $\rightarrow$ learn}, where group decisions
are made in a series of inner rounds (as in Figure \ref{fig:story-rounds})
and each outer round involves taking actions in the world.

Of note, not all of the snippets in Table \ref{tab:use-cases} involve
predictions. In particular, use case E is instructional rather than
predictive. Predictions occur or are implied in use case A (an event
is likely to occur within a defined period of time), use case B (a
thicker part G could solve the issue), use case C (children could
be harmed if action is not taken), use case D (the storyteller will
not vote for the candidate), use case F (the proposed strategy will
produce information that clarifies the situation), and use cases G
and H (the populations of rabbits and wolves will follow a certain
pattern in the future). 

\section{Story Phenomena }

Having described in broad strokes the operation of and use cases for
CogNarr apps, we now turn attention to the kinds of phenomena in natural
language (and tables and charts) that the CogNarr system should be
able to understand. The focus of this discussion is on challenges
that CogNarr natural language models face.

The first challenge is the diversity of linguistic phenomenon. A partial
categorization includes:
\begin{itemize}
\item Lexical semantics, including lexical entailment, morphological negation,
symmetry, named entities, and redundancy. 
\item Predicate-argument structure, including syntactic ambiguity, prepositional
phrases, alternations, implicits, anaphora/coreference, and intersectivity
and restrictivity. 
\item Logic, including negation, conjunction and disjunction, conditionals,
quantification, intervals and numbers, permissions, temporal structure,
and monotonicity. 
\item World knowledge and commonsense knowledge.
\end{itemize}
The Glue Benchmark provides a more complete listing of common phenomena,
along with example sentences.\footnote{\url{https://gluebenchmark.com/diagnostics} }
Linguistic phenomena\footnote{In this document, linguistic phenomena include what are commonly termed
\textit{literary} or \textit{dramatic} devices, such as metaphors
and parenthetical asides. } not explicit in the Glue Benchmark listing but of interest here include
evidentiality, tense, aspect, hyponymy, polysemy, meronymy, counterfactual
reasoning, spatial deixis, analogy, metaphor, parenthetical asides,
and approximators and other markers of uncertainty, imprecision, and
vagueness \parencite{hamawandNotionApproximation2017,fersonNaturalLanguage2015,soltVaguenessImprecision2015}. 

A second challenge is the diversity of knowledge representations \parencite{vonruedenInformedMachine2021}.
For example, stories could include, describe, or make use of: 
\begin{itemize}
\item Algebraic equations, for example $a+b<c$.
\item Differential equations, for example $\frac{du}{dt}=6u+4t$. 
\item Vectors or other structures containing numerical data, for example
$[2.3,4.9]$, as well as charts and tables. 
\item Invariances, for example spheres are invariant to rotation. 
\item Logic rules, for example $A\land B\implies C$.
\item Knowledge graphs or other kinds of graphs, for example $\underset{A}{\square}\rightarrow\underset{B}{\square}\rightarrow\underset{C}{\square}$.
\item Probabilistic relations, for example $p(a<2)=0.4$.
\end{itemize}
To give a flavor of linguistic and knowledge-representation phenomena
as they might appear in stories, selected examples from the use-case
snippets of Table \ref{tab:use-cases} are given in Table \ref{tab:linguistics}. 

Clearly, the challenges in handling a wide variety of linguistic and
knowledge-representation phenomena are substantial. No computational
language system to date can perfectly handle all the phenomena mentioned,
especially when it appears in domain-general, reasonably complex,
document-length text. Nevertheless, the goal of adequately handling
diverse phenomena in CogNarr stories is attainable. At least six characteristics
of the CogNarr ecosystem make it so:
\begin{itemize}
\item Storytellers cooperate with the system to create stories. Based on
questions and feedback provided by the system, and/or on viewing how
the system understands a passage, a user can alter a passage to accommodate
CogNarr's capabilities. For example, a user might split a long sentence
into two or more parts so that it is better understood. Or, a user
might choose a synonym for a word or multiword expression that CogNarr
is misunderstanding. Because rephrasing is possible, CogNarr does
not need to understand 100 percent of conceivable user input. It only
needs to understand a healthy majority, enough for a typical user
to adequately and naturally convey story meaning. 
\item CogNarr can restrict input to be of a form that it can process and
understand. For example, it can restrict text to adhere to a categorial
grammar formalism \parencite{steedmanSyntacticProcess2000,kohlhaseGFMMT2019}.
In this way, only certain kinds of phrases are accepted, such as noun-verb
(cats run), determiner-noun-verb (the cat ran), determiner-adjective-noun-verb
(the big cat ran), and so on. Associated rules define what constitutes
a sentence and concatenation operators specify how sentences can be
constructed from parts. Any text entry that is not a sentence according
to the formalism is rejected, perhaps along with a suggestion as to
how to write it differently so that it might be accepted. Such an
approach could also help to ensure story graph structural integrity. 
\item Groups can define domain knowledge that they wish storytellers to
use, thereby limiting the potential complexity of text. Domain knowledge
can include definitions of words, phrases, relationships, and functions,
as well as specification of commonsense knowledge bases \parencite{ilievskiDimensionsCommonsense2021}. 
\item User input can include metadata that informs the system how to process
or understand text. 
\item CogNarr is not limited to the use of only one inference model for
understanding text. Depending on the task, it can choose among models,
use hybrid models, or even combine results across multiple models. 
\item CogNarr can accept multiple kinds of knowledge representation as input,
in addition to text, and offers tools to assist with input. Even early
versions of CogNarr should be capable of reading and computing with
inputs such as equations, vectors, tables, and charts.
\end{itemize}
Taking anaphora resolution as an example, a storyteller or group can
specify character identities, characteristics, and relationships as
part of domain knowledge. This makes it easier for the system to infer
the identity of a character, given an ambiguous reference. 

Equations and functions can be entered in text or symbolic form (e.g.,
using LaTex), and certain human-generated data, such as tables and
charts, can be created using GUI tools or imported into a story graph
or fragment. For example, consider the equation of a straight line,
$y=.5x+2$, where the $x$-axis is time, in hours, and the $y$-axis
is weight, in pounds. A user can enter this equation in symbolic form
in a text box. From there, the equation can be converted into nodes
and edges of the story graph fragment (i.e., explicit representation
in the story graph format). Or the input can be converted into an
abstract syntax tree (AST) or other auxiliary representation to convey
the information. A more complicated formula or function might be converted
into a data-flow, control-flow, or other kind of graph, to provide
complementary views \parencite{bieberLibraryRepresenting2022,allamanisLearningRepresent2018}. 

Alternatively, the user might convey the equation only in text. ``There
is a function for a line. The line of interest has a slope of one-half
and intersects the $y$-axis at two.'' Or the user might enter something
more vague, such as, ``The line has a mild slope, the input ranges
between about two and eight hours, and the output ranges between about
one and ten pounds.'' Or the user might enter some observations,
``The value is about three pounds at two hours. But later on, at
about six hours, the value is five pounds.'' As yet another alternative,
the user might sketch a straight line using the sketch tool, adding
a few values to the $x$- and $y$-axis to capture its meaning. In
any of the above, the system can include uncertainty as being part
of the meaning of the input. 

In handling mathematical knowledge, a body of research can be drawn
upon that is aimed at automatically solving math and probability word
problems \parencite{falduTractableMathematical2021,zhangMultiViewReasoning2022}.
Common techniques include code generation \parencite{tangSolvingProbability2021}
and conversion to (graph-like) intermediate meaning representations
\parencite{mansouriDPRLSystems2022}. Code can also be generated from
numerical data, such as might be given in a table \parencite{blazekDeepDistilling2021}.
It is worth noting that CogNarr is not necessarily designed to solve
math and probability problems. Rather, it is designed to understand
and compute with the beliefs of storytellers. Those beliefs can, in
some cases, have exact or approximate mathematical representations. 

These are just a few examples of how CogNarr can handle or represent
linguistic and knowledge-representation phenomena. Understandably,
initial versions of CogNarr will display a limited capacity for computational
understanding of text and other input. But this capacity should increase
over time as development proceeds. What is most important is that
early versions handle enough phenomenon to be minimally useful, and
that the initial design favors later expansion toward a deeper computational
understanding of story meaning. 

\section{Related Work }

As mentioned in the Introduction, several online tools and platforms
have been developed in recent years to facilitate group deliberation
and decision making. Likewise, several projects have been developed
to aid governance of online groups or to gather and assess opinions
from online groups. This section provides a brief summary of prominent
efforts. The purpose is to contrast the features of these projects
with those of CogNarr, and to suggest ideas and sources for ideas
that could help inform or otherwise impact CogNarr development. Related
work on the topic of natural language processing and inference is
not considered here, but is discussed in the companion paper \parencite{boikCogNarrEcosystem2024b}.

\textcite{deseriisReducingBurden2023} analyzes six digital decision-making
platforms: Consul Democracy,\footnote{\url{https://consuldemocracy.org/en}}
Rousseau \parencite{deseriisDirectParliamentarianism2017,moscaDemocraticVision2020},
Polis,\footnote{\url{https://pol.is/home}} Loomio,\footnote{\url{https://www.loomio.com}}
and liquidFeedback.\footnote{\url{https://liquidfeedback.com/en}}
Each is briefly mentioned below.

Polis allows a user to create conversations and invite participants.
A conversation consists of a topic and an optional description. An
example topic from the Polis website is ``How should we protect and
restore biodiversity over the next 50 years?'' Participants can comment
on the topic using 140-character or less plain-text statements. And
they are asked to vote (agree, disagree, or pass) on the comments
submitted by others. Polis provides code that allows conversation
owners to embed the conversation in third-party websites. Polis collects
comments, any votes on comments, and for each participant, the numerical
index of the opinion group to which he or she belongs. An opinion
group might be determined, for example, via a clustering algorithm.
Polis collects representative comments from each opinion group and
the opinions most widely shared across all opinion groups. The administrative
interface displays conversation summary statistics. 

The other platforms offer variations on this theme. In Loomio, the
conversations are called \textit{threads}. Thread descriptions are
typically a few sentences to a few paragraphs in length, perhaps including
links. Thread owners can initiate a proposal or poll. A proposal can
ask participants for feedback or advice, or seek to achieve consent
or consensus on a proposal. A poll allows users to rank, score, or
otherwise indicate support. Participants can comment on a thread and
reply to comments. Comments tend to be brief.

In Consul Democracy, users can initiate or comment on proposals, debates,
participatory budgeting events, and collaborative legislation events.
Descriptions of proposals and debate topics are typically brief. Participants
can indicate their support for a proposal, and when sufficient support
is achieved, a vote can be called. Votes on proposals, debates, and
comments are yes/no. The collaborative legislation module allows a
user to view a text proposal, make comments, and vote on the comments
of others.

LiquidFeedback allows a user to create an initiative, indicate support
for an initiative, and suggest improvements to an initiative. Descriptions
of initiatives tend to be brief. Votes are either yes/no or a set
of preferences for first choice, second choice, and so on. Users can
delegate their vote to others. 

Although not reviewed by \textcite{deseriisReducingBurden2023}, Decidim\footnote{\url{https://decidim.org}}
is another project of interest. It offers features that are similar
to the others mentioned, and allows for blog posts, the creation of
participatory texts, and the establishment of permanent assemblies,
such as councils and commissions. 

In comparison to these frameworks, CogNarr offers a unique set of
features that are more formally rooted in the cognitive process, as
it is currently understood by science. This includes communication
about beliefs, expectations, and uncertainties. Stories can be rich,
document length, and involve a wide range of linguistic and knowledge-representation
phenomena. A computational system aids in analysis and inference.
The system can read, understand, answer questions about, and compute
with the set of stories created by a potentially large group. Moreover,
the system can offer real-time feedback and ask questions to storytellers,
to help them improve stories. The CogNarr ecosystem can be employed
in a wide range of use cases, not all of which are directly aimed
at group deliberation and decision making. 

In his paper, \textcite{deseriisReducingBurden2023} focuses on how
frameworks implement political equality in three stages of the decision-making
process: agenda setting, voting, and exchange and synthesis of opinions.
Political equality and the three processes are also important in the
CogNarr ecosystem. The first two, agenda setting and voting, are handled
at the group level. Groups have flexibility as to how they operate,
and there are numerous procedural options they can choose from. The
particulars of groups, such as their organization, procedures, and
rules, are transparent to members. 

The third process, exchange and synthesis of opinions, is central
to CogNarr operations, especially for events aimed at decision making,
and especially when editing occurs over a series of rounds (as illustrated
in Figure \ref{fig:story-rounds}). Here, the exchange and synthesis
of opinions occurs through the sharing of stories and through user-driven
edits to stories, which are often made in response to feedback, and
after viewing the stories of others. 

Political equality is favored, as each person submits only one story
(unless group rules allow otherwise), all stories are assessed via
the same transparent process, and results are summarized and presented.
In this way, no voice is hidden or filtered, and no voice is necessarily
bigger than others, except as influenced by reputation and story quality.
A deeper exploration of political equality and related topics is left
for future work.

Another effort of note is the Metagovernance Project,\footnote{\url{https://metagov.org}}
a research collective focused on building standards and infrastructure
for digital self-governance. Some Metagovernance projects appear to
be a good match for the CogNarr ecosystem. For example, PolicyKit
\parencite{zhangPolicyKitBuilding2020,wangPikaEmpowering2023}, CommunityRule,\footnote{\url{https://communityrule.info}}
and ModularPolitics \parencite{schneiderModularPolitics2021} help
online communities to develop and implement governance processes and
rules. A CogNarr group could conceivably use these resources to help
define how it operates and the procedures by which it makes decisions. 

Deliberative Democracy Lab\footnote{\url{https://deliberation.stanford.edu}}
is another project of note aimed at group deliberation. Their programs
include Deliberative Polling, in which a sizable and diverse group
of people is selected to engage in online video discussions about
a topic. The detailed opinions of the group are recorded after discussions
conclude.

\textcite{shortallReasonMachine2022} provide a literature review
on design features of online deliberation platforms, with focus on
the challenges of scaling deliberation to large crowds. Topics include
quality of the deliberation process and inclusion of diverse social
groups. These are also important topics in CogNarr, but addressing
them is left for future work. 

Lastly, there is a sizable body of work on large-group decision making
\parencite{garcia-zamoraLargeScaleGroup2022,tangConventionalGroup2021}.
The focus of this work tends to be on mathematical approaches that
identify, weigh, and summarize preference differences in a group,
and optimal choices for decision makers given preference differences
among stakeholders. Some of these ideas could be used by CogNarr to
identify and summarize preference differences within a group, during
rounds of story evolution. 

\section{Conclusion}

The CogNarr project is ambitious, and clearly requires a team effort
to develop the ideas and infrastructure to the point where a minimum
viable product becomes available. The project is complex and success
will require considerable effort. To date, this paper and its companion
are the first and only steps taken. It is my hope that the two papers
generate interest in and questions about the CogNarr project. Whether
or not interest grows might depend, in part, on whether others are
inspired by the vision presented. To that end, I now recap and expand
upon some potential benefits to society. 

A few potential benefits were already mentioned in the Introduction---an
improved capacity to address and solve problems, prevent and recover
from harm, and identify and act on shared purpose. If CogNarr were
successful, and if in time these benefits were actualized, the positive
impacts on the health and wellbeing of societies, and their ecosystems,
might be profound. Any number of social, financial, environmental,
commercial, technical, or other types of problems might be successfully
addressed, or at least addressed more successfully than is now common
or possible. A successful CogNarr ecosystem might also inspire innovations
within the governance systems used by political entities. These prospects
alone might be reason enough to develop the CogNarr project. But there
is another reason. 

Efforts to improve group cognition might deeply resonate with who
we are as humans. That is, our desire for functional group cognition
might reflect a very deep and ancient desire, ingrained in us through
millennia of evolution. As \textcite{lyonCognitiveCell2015} and others
point out, cognition is not restricted to humans; it is a universal
characteristic of life. It has been present since life first appeared
on the planet. Among other things, cognition is central to the cooperation
evident throughout the biological kingdom. Even more, cognition rests
on learning, and learning is what the universe appears to be doing
all around us \parencite{vanchurinTheoryEvolution2022}. Learning---construed
abstractly as an optimization process, or minimization of a loss function---is
not just a characteristic of life, it appears to be a characteristic
of everything that exists, animate and inanimate, from galaxies to
atoms. In this view, the creation and evolution of biological life,
and its display of cognition, is but one, wondrous, example. Thus,
by improving group cognition, CogNarr seeks to better align human
activity with the process that surrounds and infuses us. One might
deduce that the very purpose of humans, gifted to us by the universe,
is to learn, even to excel at learning. More broadly, our purpose
might be to support, express, and expand intelligence, within and
without. This would explain our innate desire to explore and understand,
to act with wisdom, to value nature, and to pass on what is learned
to new generations. 

We can fail to fulfill this deep purpose. We can fail to learn from
our mistakes, fail to value what is valuable, and fail to adapt to
changing conditions. In that case, our time as a species, or at least
our time as a flourishing species, might be limited. Or, we can fulfill
our purpose unconsciously---slowly, painfully, by stumbling in fits
and starts. As another alternative, we can fulfill our purpose consciously,
with gusto and caring. Which path we take is up to us. I for one hope
that we choose the conscious path. Importantly, that path is not simply
about behavioral change---acting more thoughtfully, for example,
listening more carefully, or allowing for more curiosity. It is also
about reorganizing our social processes, institutions, rules, norms,
and systems until they faithfully align with our deep purpose. The
closer that alignment, the better we serve ourselves and the greater
good, and the more beautiful the world becomes. Hence, the CogNarr
project. It represents a reorganization of, and new tools for, the
process by which human groups consciously and purposefully cognate.

\medskip{}

\textbf{Acknowledgment:} I would like to thank Lee Eyre and Tim Lloyd,
who over lengthy conversations helped develop the initial CogNarr
concept. I would also like to thank Lee, Tim, Daniel Friedman, Avel
Guenin-Carlut, and Mahault Albarracin for offering comments on drafts
of this paper.

\medskip{}

\printbibliography

\end{document}